\documentclass[runningheads]{llncs}

\makeatletter                   %1
\def\mdseries@tt{m}             %1
\makeatother                    %1

\usepackage[draft=true]{minted} %2

\usepackage{graphicx}
\usepackage{amsfonts}
\usepackage{amssymb,amsbsy}
\usepackage{amsmath}
\usepackage{amsfonts, amscd}
\usepackage{algorithm, algorithmic}
\usepackage{color}
\usepackage{url}
\usepackage{verbatim}
\usepackage{setspace}
\usepackage[abs]{overpic}
\usepackage{graphicx}
\usepackage{subfigure}
\usepackage{bbm, bm}
\usepackage{dsfont}
\usepackage{bbm}
\newcommand{\mb}{\boldsymbol}
\renewcommand{\bm}{\boldsymbol}
\newcommand{\mc}{\mathcal}

\newcommand{\reals}{\mathbb{R}}
\newcommand{\<}{\langle}
\renewcommand{\>}{\rangle}
\newcommand{\innerprod}[2]{\left\< #1, #2 \right\>}
\newcommand{\set}[1]{\left\{ #1 \right\}}

\newlength{\dhatheight}

\newcommand{\diag}[1]{\mathrm{diag}\left( #1 \right)}

\usepackage{booktabs} % For formal tables
\usepackage{multirow}
\usepackage{array}
\usepackage{multicol}
\usepackage{minted}
\usepackage{arydshln}

\newcolumntype{L}[1]{>{\raggedright\let\newline\\\arraybackslash\hspace{0pt}}m{#1}}
\newcolumntype{C}[1]{>{\centering\let\newline\\\arraybackslash\hspace{0pt}}m{#1}}
\newcolumntype{R}[1]{>{\raggedleft\let\newline\\\arraybackslash\hspace{0pt}}m{#1}}

\begin{document}
\title{Fast and Exact Nearest Neighbor Search in Hamming Space on Full-Text Search Engines}
\titlerunning{FENSHSES}
% If the paper title is too long for the running head, you can set
% an abbreviated paper title here
%
\author{Cun (Matthew) Mu\inst{1}\and
Jun (Raymond) Zhao \inst{1}\and
Guang Yang \inst{1}\and \\
Binwei Yang\inst{2}\and
Zheng (John) Yan\inst{1}}
\authorrunning{C. Mu et al.}
% First names are abbreviated in the running head.
% If there are more than two authors, 'et al.' is used.
%
\institute{Jet.com/Walmart Labs, Hoboken, NJ \\
\email{\{matthew.mu, raymond, guang, john\}@jet.com}\\
\and
Walmart Labs, Sunnyvale, CA\\
\email{BYang@walmartlabs.com}}
\maketitle              % typeset the header of the contribution
\begin{abstract}
	A growing interest has been witnessed recently from both academia and industry in building nearest neighbor search (NNS) solutions on top of full-text search engines. Compared with other NNS systems, such solutions are capable of effectively reducing main memory consumption, coherently supporting multi-model search and being immediately ready for production deployment. In this paper, we continue the journey to explore specifically how to empower full-text search engines with fast and exact NNS in Hamming space (i.e., the set of binary codes). By revisiting three techniques (bit operation, subs-code filtering and data preprocessing with permutation) in information retrieval literature, we develop a novel engineering solution for full-text search engines to efficiently accomplish this special but important NNS task. In the experiment, we show that our proposed approach enables full-text search engines to achieve significant speed-ups over its state-of-the-art term match approach for NNS within binary codes. 
%	
%	This achieves a coherent approach to multi-modal searches (e.g., allowing customers to express their interests in both visual and textual requests), in which traditional nearest neighbor search systems are incompetent. In this paper, we focus specifically on nearest neighbor search within binary codes, which are commonly used to semantically represent documents (e.g., images, sounds and videos). By revisiting three techniques in traditional nearest neighbor search systems: bit operation, substring filtering and data preprocessing with permutation, we end up with a novel approach to achieving nearest neighbor search in Hamming space, which achieves dramatic speed-ups over the existing term match baseline.
	
%	within Elasticsearch--one of the most popular full-text search engines. In this paper, we focus specifically on Hamming space nearest neighbor search using Elasticsearch. By combining three techniques: bit operation, substring filtering and data preprocessing with permutation, we develop a novel approach called FENSHSES (Fast Exact Neighbor Search in Hamming Space on Elasticsearch), which achieves dramatic speed-ups over the existing term match baseline. This will empower Elasticsearch with the capability of fast information retrieval even when documents (e.g., texts, images and sounds) are represented with binary codes--a common practice in nowadays semantic representation learning.
	

\end{abstract}
\section{Introduction}\label{sec: intro}
Full-text search engines, based on first-order document-term statistics such
as TF-IDF and Okapi BM25, have been deployed ubiquitously in nowadays web applications to help customers find textual documents that match their specified keywords. 

Recently, active efforts from both academia and industry \cite{lux2013visual,rygl2017semantic,ruuvzivckaflexible,amato2018large,mu2018towards} have been witnessed to empower full-text search engines with the capability of nearest neighbor search (NNS). Compared with other NNS solutions (e.g., Annoy \cite{Github:annoy}, FLANN \cite{muja2014scalable}, FAISS \cite{johnson2017billion} and  SPTAG  \cite{ChenW18}),  such full-text search engine based ones have a number of clear advantages.

\paragraph{Implemented in secondary memory.} As demonstrated by Amato et al. (2018), unlike other NNS solutions implemented in main memory , due to the highly optimized disk-based index mechanics behind full-text search engines, NNS systems established on full-text search engines substantially reduce main-memory consumption. This makes such systems more cost-effective and thus more suitable to big-data applications. 

\begin{figure}[t]
	\centerline{\includegraphics[width=4in]{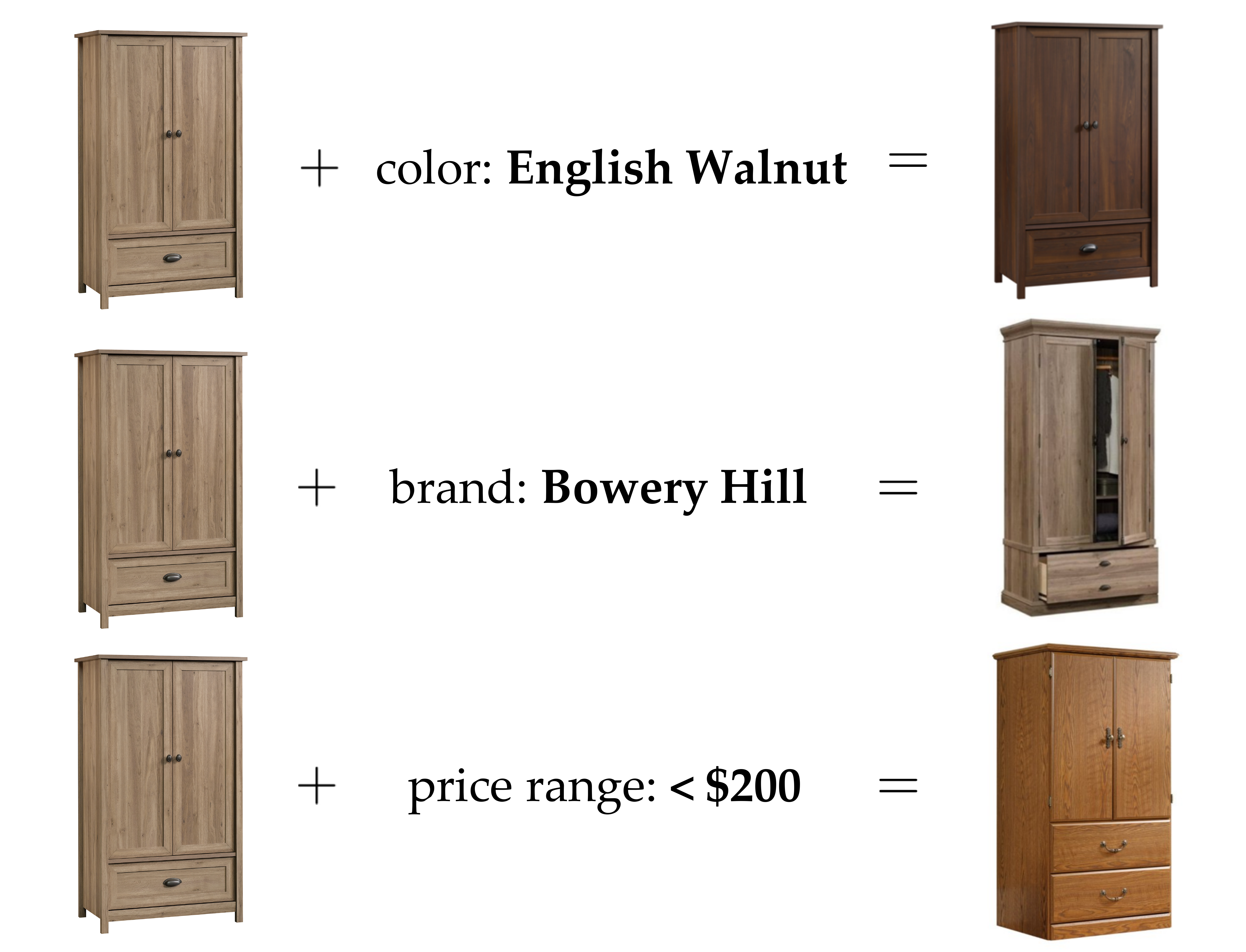}}
	\caption{{\bf Illustration of multi-modal search.} \normalfont{Full-text search engines, empowered with nearest neighbor search, allow customers to express their interests simultaneously in image queries (whose visual feature vectors will be consumed in NNS to find visually similar products) and textual queries (e.g., color, brand and price range). As a result, products retrieved will not only be visually similar to uploaded images but also satisfy requested keywords from customers.}} \label{fig: ms}
\end{figure}

\paragraph{Flexible in multi-model search.} As highlighted by Mu et al. (2018), enabling full-text search engines with NNS paves a coherent way for multi-model searches--e.g., allowing users to express their interests in both visual and textual queries (see Fig. \ref{fig: ms} for an application of multi-model search in eCommerce)--at which most of other NNS systems fall short. 
 
\paragraph{Ready for production.} Last but not least, as emphasized by Rygl et al. (2017), NNS systems built upon full-text search engines are extremely well-prepared for production deployment. Due to the cutting-edge engineering designs from full-text search engines (e.g., Elasticsearch and Solr), important features like horizontal scaling, I/O and cache optimization, security configuration, index and cluster management, real-time monitoring and RESTful APIs are immediately ready to be consumed by such NNS systems, so that engineers can effectively avoid reinventing the wheel themselves. 

Blessed by all the above distinct benefits, we continue this journey to explore specifically effective ways to achieve {exact nearest neighbor search in Hamming space (i.e., the set of binary codes) on top of full-text search engines.}
%\footnote{Note that even though our discussions and implementations are pertained with Elasticsearch, the methodology introduced in the paper is generally applicable to any full-text search engine (e.g, Solr and Sphinx).} 
\paragraph{Problem statement.} Specifically, with the following dataset of binary codes, 
\begin{flalign}
\mc B = \set{\bm b_1, \bm b_2, \ldots, \bm b_n} \subset \set{0,1}^m,
\end{flalign}
the goal of our paper is to enable full-text search engines with the capability of efficiently finding all $r$-neighbors of $\bm q$ in $\mc B$, namely
\begin{flalign}
B_H(\bm q, r) := \set{\bm b \in \mc B \;\vert\; d_H(\bm b, \bm q) \le r},
\end{flalign}
where $d_H(\bm b, \bm q):=\sum_{i=1}^m  \mathbbm 1_{\set{b_i \neq q_i}}$ denotes the Hamming distance between binary code $\bm b$ and $\bm q$.\footnote{It is worth noting that the $r$-neighbor search problem studied by the paper can also be easily adapted to conduct $k$-NN ($k$-nearest neighbors) search by progressively increasing the Hamming search radius $r$ until $k$ neighbors are found.} 

\paragraph{Why binary codes?} Finding nearest neighbors in Hamming space is an extremely important subclass of NNS, as learning and representing textual or visual data with compact and semantic binary vectors is a pretty mature technology and common practice in nowadays information retrieval. Using well-trained binary vectors instead of floating ones enables dramatic reductions in storage and communication costs without too much sacrifice in search accuracy. In particular, eBay builds its whole visual search system \cite{yang2017visual}  by finding nearest neighbors within binary codes extracted from catalog and query images through the supervised semantic-preserving deep Hashing (SSDH) model \cite{lin2015deep}.  

\paragraph{What is missing?}  However, most of the aforementioned full-text search engines based NNS solutions \cite{rygl2017semantic,ruuvzivckaflexible,amato2018large,mu2018towards} would fail to deal with binary codes, as they find nearest neighbors on full-text search engines by generating and indexing surrogate textual tokens merely based on information collected from several top entries of each floating vector in terms of magnitude. The only exception is the term match approach developed by Lux and Marques (2013) in their Java library called  LIRE (Lucene Image Retrieval). The core idea is to calculate Hamming distance between two binary codes by matching their bits at each position. This term match approach on one hand is a natural way to leverage full-text search engines to conduct NNS, but on the other hand heavily overlooks the intrinsic and special properties within binary codes by treating binary digits simply as texts. Motivated by this, we revisit three pervasive techniques from information retrieval literature: bit operation, sub-code filtering and data preprocessing with permutation. By integrating these three techniques seamlessly into full-text search engines, we end up with a solution that outperforms the term match one dramatically in terms of search latency. 

\paragraph{Elasticsearch.} Elasticsearch (ES), built upon Apache Lucene, is an open-source, real-time, distributed and multi-tenant full-text search engine. Since its first release in Feb. 2010, it has become the most popular enterprise search engine and widely adopted by a variety of companies (e.g., Ebay, Facebook, GitHub, Lyft and Shopify) for either internal or external uses to discover relevant documents. Similar to previous works \cite{rygl2017semantic,ruuvzivckaflexible,amato2018large,mu2018towards}, without the loss of general applicability to other full-text search engines, we elaborate our core ideas concretely  on the platform of Elasticsearch. First-hand experiences in implementing this ES-based solution are extensively addressed, which should be greatly valuable for practitioners to understand and replicate our approach. 

\paragraph{Organization.} The rest of the paper is organized as follows. In Section \ref{sec: term_match}, we first review the term match approach widely implemented on full-text search engines to find nearest neighbors among binary codes. 
In Section \ref{sec: fenshses}, we propose a better one for full-text search engines to accomplish this task. Specifically, we implement an Elasticsearch-based solution called FENSHSES (Fast Exact Neighbor Search in Hamming Space on Elasticsearch) to conduct nearest neighbor search in Hamming space. We incorporate three techniques into FENSHSES: {\em bit operation}, which enables Elasticsearch to compute Hamming distance with just a few bit operations; {\em sub-code filtering}, which instructs Elasticsearch to conduct a simple but effective screening process before any Hamming distance calculation and therefore empower FENSHSES with sub-linear search times; {\em data preprocessing with permutation}, which preprocesses
 binary codes with appropriate permutation to maximize the effect of sub-code filtering. In Sec. \ref{sec: exp}, we show that FENSHSES  outperforms the term match approach dramatically in terms of search latency.

%However, most  of the aforementioned scalable NNS solutions implemented on ES conduct approximate NNS by generating and indexing surrogate textual tokens merely based on information collected from several top entries of each floating vector in terms of magnitude, which would clearly fail for this Hamming case. Currently, the widely used approach on ES for exact Hamming space NNS is the term match one from LIRE, which we will review in Sec. \ref{sec: term_match}. Its core idea is to calculate Hamming distance between two binary codes by matching their bits at each position. This is a natural way to leverage ES as a full-text search engine. However, this term approach treats binary digits in a textual way and heavily overlooks the intrinsic and special properties within binary codes. Motivated by this, in Sec. \ref{sec: fenshses}, we develop a novel approach called  {\em Fast Exact Neighbor Search in Hamming Space on Elasticsearch (FENSHSES)} by combining three techniques: {\em bit operation}, which enables Elasticsearch to compute Hamming distance with just a few bit operations; {\em sub-code filtering}, which instructs Elasticsearch to conduct a simple but effective screening process before any Hamming distance calculation and therefore empower FENSHSES with sub-linear search times; {\em data preprocessing with permutation}, which preprocesses binary codes with appropriate permutation to maximize the effect of sub-code filtering. In Sec. \ref{sec: exp}, we show that FENSHSES  outperforms the term match baseline dramatically in terms of search latency.

\section{Term Match from LIRE}\label{sec: term_match}
Based on its definition, Hamming distance is nothing but the number of positions at which two binary codes vary.  As a result,  full-text search engines can naturally compute this through term match. Specifically, for each binary code $\bm b$, we can index its positions corresponding to ones and zeros, respectively, i.e., 
\begin{flalign}
\mc O_b = \set{i \in [m] \;\vert\; b_i = 0} \;\mbox{and } \; \mc I_b = \set{i \in [m] \;\vert\; b_i = 1}, \nonumber
\end{flalign}
with $[m]:=\set{1,2,\ldots, m}$. When the query binary code $\bm q$ arrives, Elasticsearch can simply calculate its Hamming distance to each binary code $\bm b \in \mc B$ by matching its zero and one positions with $\bm b's$:
\begin{flalign}\label{eqn: tm}
d_H(\bm b, \bm q) = m -\left( \sum_{i \in \mc I_q} \mathbbm 1_{\set{i \in \mc I_b}} + \sum_{j \in \mc O_q} \mathbbm 1_{\set{j \in \mc O_b}}\right).
\end{flalign}
The JSON-encoded request body for Elasticsearch to find $B_H(\bm q, r)$ by computing \eqref{eqn: tm} is illustrated in JSON \ref{json: term_match},\footnote{All ES-related implementations are based on Elasticsearch 6.1.0.} where we denote $u:=|\mc I_q|$ and $v:=|\mc O_q|$. 

This term match approach, firstly introduced in the Java library LIRE \cite{lux2013visual} to find visually similar images (based on their binary visual features), is currently the cutting-edge approach for full-text search engines to find nearest neighbors within binary codes. Some of its variants (e.g., using fuzzy query based on Levenshtein edit distance)  are also widely used on  full-text search engines nowadays. 

\AtBeginEnvironment{minted}{%
	\renewcommand{\fcolorbox}[4][]{#4}}
\renewcommand{\listingscaption}{JSON}
\begin{listing}[h]
{
\begin{minted}[linenos=true]{json}
{
  "min_score": n-r,
  "query": {
    "function_score": {
      "functions":[
        {"filter": {"term": {"Ib": Iq[1]}}, "weight": 1},
        ...,
        {"filter": {"term": {"Ib": Iq[u]}}, "weight": 1},
        {"filter": {"term": {"Ob": Oq[1]}}, "weight": 1},
        ...,
        {"filter": {"term": {"Ob": Oq[v]}}, "weight": 1}
      ],
      "score_mode": "sum",
      "boost_mode": "replace"
    }
  }
}  
\end{minted}
 }
\caption{ES request body of the term match approach to  $r$-neighbor search in Hamming space}
\label{json: term_match}
\end{listing}

\section{Proposed Approach: FENSHSES}\label{sec: fenshses}
The term match approach treats each binary digit (i.e., bit) in a textual way, which heavily overlooks the intrinsic and special properties of binary codes. By making better uses of these properties, we introduce a novel approach called FENSHSES (Fast Exact Neighbor Search in Hamming Space on Elasticsearch), whose complete JSON-encoded ES request body can be found in JSON \ref{json: fenshses}. In essence, FENSHSES integrates three techniques: bit operation, sub-code filtering and data preprocessing with permutation, which should be generally applicable to other full-text search engines besides Elasticsearch. These three techniques are pervasively used in nearest neighbors search for binary codes; but as fas as we know, this is the first-time such techniques are seamlessly integrated into full-text search engines, and thus leads to a novel NNS solution with minimal main memory consumption, full support in multi-modal search and extreme readiness to be deployed in production (per our discussions in Section \ref{sec: intro}).

\subsection{Bit Operation}
Motivated by the well-known fact that hamming distances between binary codes can be computed extremely fast using bit operations, in this part, we will explore how we can replace term match by natively empowering Elasticsearch to calculate hamming distances through bit operations. 

For an $m$-bit binary code $\bm b$, we will first segment it into $s$ sub-codes:\footnote{For simplicity, we assume $s$ divides $m$.} 
\begin{flalign}\label{eqn: subcodes}
[\underbrace{b_1, \ldots, b_{\frac{m}{s}}}_{\bm b^1}, \underbrace{b_{\frac{m}{s}+1}, \ldots, b_{\frac{2m}{s}}}_{\bm b^2}, \ldots \ldots, \underbrace{b_{m-\frac{m}{s}+1}, \ldots, b_{m}}_{\bm b^s}].   
\end{flalign}
Since $d_H(\bm q, \bm b) = \sum_{i\in [s]} d_H(\bm q^i, \bm b^i) $, the Hamming distance calculation is reduced into $s$ ones with binary codes of much shorter length. We re-implement the assembly codes found in the notable HAKMEM memo \cite{beeler1972hakmem} to compute the Hamming distance between two short binary codes of length 64 or less into {Painless}--a simple and secure scripting language designed specifically for Elasticsearch. When the query binary code $\bm q$ is issued, we will invoke hmd64bit $s$ times to calculate $\set{d_H(\bm q^i, \bm b^i)}_{i=1}^s$ by specifying $\bm q^i$ and $\bm b^i$  as parameters accordingly and then sum them up. The whole process can be efficiently implemented in ES using the {\em function score query}, where several functions are combined to calculate the score of each document (see lines 15-31 in JSON \ref{json: fenshses}).

\begin{listing}[h]
{
\begin{minted}[linenos=true]{json}
POST _scripts/hmd64bit
{
  "script": {
    "lang": "painless",
    "source": """
      long u = params.subcode^doc[params.field].value;
      long uCount = u-((u>>>1)&-5270498306774157605L) 
                               -((u>>>2)&-7905747460161236407L);
      return ((uCount+(uCount>>>3))&8198552921648689607L)%63;
    """
  }
}
\end{minted}
}
\caption{Create the script called hmd64bit into Elasticsearch through the {\em \_scripts} end-point.}
\label{json: hmd64bit}
\end{listing}

\subsection{Sub-Code Filtering}
So far, regardless of the term match approach or the bit operation one, we have to exhaustively compute the Hamming distance between $\bm q$ and each binary code in $\mc B$. This expensive linear scan is not desirable for many applications where the number of codes in $\mc B$ is in the order of millions or even billions \cite{yang2017visual}. As a remedy, in this part, we will borrow a simple but powerful counting argument from Norouzi et al. (2012) to conduct a screening process before any Hamming distance calculation, which successfully empowers our FENSHSES approach with sub-linear search times. 

Suppose binary codes are segmented into $s$ sub-codes as in \eqref{eqn: subcodes}. Then for two codes $\bm b$ and $\bm q$ within $r$ Hamming distance, among all their $s$ sub-code pairs $\set{(\bm b^i, \bm q^i)}_{i=1}^s$, there must be at least one pair with Hamming distance no larger than $\lfloor \frac{r}{s} \rfloor$, which mathematically implies 
\begin{flalign}\label{eqn: substring-filter}
B_H(\bm q, r) \subseteq \bigcup_{i=1}^s \set{\bm b \in \mc B \; \left \vert  \;  \bm b^i \in B_H\left(\bm q^i, \lfloor \frac{r}{s} \rfloor \right.\right)}. 
\end{flalign}
%This simple counting argument, firstly leveraged by Norouzi et al. (2012) to build multiple hash tables for fast search in Hamming space, yields great potentials in reducing the number of Hamming distance calculations needed to find all $r$-neighbors $\bm q$ in $\mc B$. Specifically, according to relationship \eqref{eqn: substring-filter}, it is safe to just consider binary codes belonging to the set on the right side of \eqref{eqn: substring-filter}, whose size could be substantially smaller than $n$  for $r \ll m$. 
This simple counting argument yields great potentials in reducing the number of Hamming distance calculations needed to find all $r$-neighbors $\bm q$ in $\mc B$. Specifically, according to relationship \eqref{eqn: substring-filter}, it is safe to just consider binary codes belonging to the set on the right side of \eqref{eqn: substring-filter}, whose size could be substantially smaller than $n$  for $r \ll m$. It is worth noting that similar ideas have been frequently revisited in many different contexts--e.g., multi-index hashing \cite{norouzi2012fast} and  string similarity joins \cite{li2012pass}, and a generalized version of \eqref{eqn: substring-filter} is also derived  recently \cite{qin2018gph}.

Due to the inverted-indexing nature of full-text search engines, this sub-code filtering step is extremely suitable and straightforward to be implemented on full-text search engines. Specifically, on Elasticsearch, we can simply leverage the {\em filter context} (see lines 8-14 in JSON \ref{json: fenshses}), within which each sub-code Hamming ball $B_H\left(\bm q_i, \lfloor \frac{r}{s} \rfloor \right)$ is obtained by the {\em terms query} (e.g., line 11 in JSON \ref{json: fenshses}), and the union is achieved through a {\em boolean combination of should clauses}.

%It is worth noting that similar ideas have been frequently revisited in many different contexts--e.g., to build multi-index hashing tables \cite{greene1994multi, norouzi2012fast}.

\subsection{Data Preprocessing with Permutation}
The effectiveness of sub-code filtering will be maximized if the bits within the same sub-code group are statistically independent. Since hamming distance is invariant to permutation transformation, it is tempting to transform binary codes in $\mc B$ with appropriate permutation towards this desired group independence property.  

For two Bernoulli random variables $x$ and $y$, they are independent if and only if their correlation coefficient $\rho(x,y)=0$. Therefore, it is natural to find a permutation $\bar \pi$ to minimize correlation effects among each sub-code segment. This immediately leads to the following optimization problem proposed by Wan et al. (2013) to improve the performance of \cite{norouzi2012fast}:
\begin{flalign}
\min_{\pi: [m] \to [m]} &	 \quad \innerprod{\mb D}{\mb P_\pi \mb M_{\mc B} \mb P_{\pi}^\top} \nonumber\\
\quad	\mbox{s.t.} \; \quad  &	\quad   \pi \mbox{ is a permutation}. \label{eqn: bgp}
\end{flalign}
Here $\mb D = \diag{\mb I_{d\times d}, \ldots, \mb I_{d\times d}} \in \reals^{m \times m}$ is a block diagonal matrix with $\mb I_{d\times d}$ as a matrix of ones and $d = m/s$,  $\mb P_{\pi}$ is the permutation matrix induced by $\pi$:
\begin{flalign} \nonumber
\mb P_{\pi} = 
\begin{bmatrix}
\bm e_{\pi(1)} & \bm e_{\pi(2)} & \cdots & \bm e_{\pi(m)}
\end{bmatrix}^T
\end{flalign}
and $\mb M$ is a matrix in $ \reals^{m \times m}$ whose $(i,j)$-entry is obtained from $\mc B$ as the absolute value of the correlation between the $i$-th and the $j$-th bits.

Problem \eqref{eqn: bgp} is essentially an instance of the extensively studied balanced graph partition problem \cite{kernighan1970efficient,andreev2006balanced,krauthgamer2009partitioning,bader2013graph}. In FENSHSES, we 
solve problem \eqref{eqn: bgp}  by the well-known and scalable Kernighan-Lin algorithm \cite{kernighan1970efficient}, the gist of which is to find appropriate pairs $i,j \in [m]$ and swap their mappings in $\pi$ with 
\begin{flalign}
\left(\pi(i), \pi(j)\right) \gets \left(\pi(j), \pi(i)\right). \nonumber
\end{flalign}
We leave it as a future work to solve \eqref{eqn: bgp} by more recently developed approximation algorithms with better theoretical guarantees--e.g.,  the one based on semidefinite relaxation \cite{krauthgamer2009partitioning}.
%\vspace{-5mm}
%\begin{flalign} \nonumber
%\left(\pi(i), \pi(j)\right) \;\gets\;\left(\pi(j), \pi(i)\right).
%\end{flalign}

%and then preprocess binary codes in $\bm b \in \mc B$ with $\pi^\star$, i.e., 
%%\begin{flalign}
%%	\mc B^{\pi^\star}  := \set{\mb P^{\bm \pi^\star} \bm b \; \vert \; \bm b \in \mc B}, 
%%\end{flalign}
%\begin{flalign}
% \bm b^{\bm \pi^\star}:= \mb P^{\bm \pi^\star} \bm b = 	\begin{bmatrix}
% 	b_{\pi^\star(1)} & b_{\pi^\star(2)} & \cdots & b_{\pi^\star(m)}
% \end{bmatrix}^T,
%\end{flalign}
%where $\mb P^{\pi^\star}$ is the permutation matrix induced by $\pi^\star$ with
%\begin{flalign} \nonumber
%	\mb P^{\pi^\star} = 
%	\begin{bmatrix}
%		\bm e_{\pi^\star(1)} & \bm e_{\pi^\star(2)} & \cdots & \bm e_{\pi^\star(m)}
%	\end{bmatrix}^T.
%\end{flalign}
%This naturally leads to the following mathematical optimization proposed by Wan et al. \cite{wan2013data}:
%\begin{flalign}
%\pi^\star \in	\min_{\pi: [m] \to [m]} &	 \quad \innerprod{\mb D}{\mb P_\pi \mb M \mb P_{\pi}^\top} \nonumber\\
%\quad	\mbox{s.t.} \; \quad  &	\quad   \pi \mbox{ is a permutation}.
%\end{flalign}
%Here $\mb D = \diag{\mb I_{d\times d}, \ldots, \mb I_{d\times d}} \in \reals^{m \times m}$ is a block diagonal matrix with $\mb I_{d\times d}$ as a matrix of ones and $d = m/s$,  and  $\mb M$ is a matrix in $ \reals^{m \times m}$ whose $(i,j)$-entry is estimated from $\mc B$ as the absolute value of the correlation between the $i$-th and the $j$-th bits.

\subsection{Elasticsearch index for FENSHSES}
For the proposed FENSHSES approach, the data to be indexed into Elasticsearch cluster is pretty minimal. For each binary code $\bm b \in \mc B$, we only need to index its sub-codes $\set{\bm b^1, \bm b^2, \ldots, \bm b^s}$.\footnote{More accurately speaking, what being indexed are the integers represented by the binary sub-codes.}  In JSON \ref{json: fenshses_index}, we demonstrate how to index sub-codes together with other fields into one Elasticsearch index. Here we assume that bit operation and sub-code filtering segment binary codes in the same manner. If their segmentations are different, it might be recommended to index two sets of sub-codes into the same Elasticsearch index by treating each set as a {\em nested datatype}. 

Suppose sub-codes $\set{\bm b^1, \bm b^2, \ldots, \bm b^s}$ are semantic visual embeddings from product images. With other product information (e.g., title, brand and price) indexed together with $\set{\bm b^1, \bm b^2, \ldots, \bm b^s}$  into one Elasticsearch index, we can easily achieve multi-model searches depicted in Figure  \ref{fig: ms} by adding filters (e.g., term filter and range filter) into JSON \ref{json: fenshses}.

\begin{listing}[h]
{
\begin{minted}[linenos=true]{json}
PUT /fenshses
PUT /fenshses/default/_mapping
{
"properties": {
  "title": {"type": "text"},
  "brand": {"type": "keyword"},
  "price": {"type": "double"},
  "is_in_stock": {"type": "boolean"},
  ...,
  "b1": {"type": "long"}, 
  ..., 
  "bs": {"type": "long"}
  }
}
\end{minted}
}
\caption{Request body to create the Elasticsearch index and define its mapping to support the FENSHSES approach.}
\label{json: fenshses_index}
\end{listing}

\AtBeginEnvironment{minted}{%
	\renewcommand{\fcolorbox}[4][]{#4}}
\renewcommand{\listingscaption}{JSON}
%\setminted{mathescape, escapeinside=||}
\begin{listing}[hb]
{
\begin{minted}[linenos=true]{json}
{
  "min_score": m-r,
  "query": {
    "function_score":{
      "query":{
        "constant_score":{
          "boost": m,
          "filter":{
            "bool":{
              "should":[
                {"terms":{"b1": r/m-neighbor of q1}},
                ...,
                {"terms":{"bm": r/m-neighbor of qs}}
              ]
            }
          }
        }
      }, 
      "functions":[
        {"script_score": {"script": 
                           {"id": "hmd64bit", 
                            "params": {"field": "b1", "subcode": q1}}},
          "weight": -1},
        ...
        {"script_score": {"script": 
                           {"id": "hmd64bit", 
                            "params": {"field": "bm", "subcode": qm}}},
          "weight": -1}
        ],
        "boost_mode": "sum",
        "score_mode": "sum"
    }
  }
}
\end{minted}
}
\caption{Elasticsearch request body of the FENSHSES approach to $r$-neighbor search in Hamming space}
\label{json: fenshses}
\end{listing}

\section{Experiment}\label{sec: exp}
We compare search latencies between the term match approach and FENSHSES with semantic binary codes generated from Jet.com's catalog images. To better understand the contribution of each technique involved in FENSHSES, we experiment systematically with  four methods: the term match baseline, FENSHSES with just bit operation, FENSHSES without data preprocessing and FENSHSES.

\paragraph{Settings.} Our dataset $\mc B$ is generated using half a million images selected from Jet.com's furniture catalog through the SSDH model \cite{lin2015deep}, which leverages deep CNN (convolutional neural network) to output semantic binary codes in an end-to-end manner\footnote{Note that the purpose of the experiment is not to compare different embedding models, but to evaluate the performance of FENSHESES, which should be generally applicable to NNS in any Hamming space.}. We choose  the length of binary codes to be $128$ and $256$ respectively. For the setting of FENSHSES, we keep the sub-code length as $64$ for bit operation and $16$ for sub-code filtering throughout the experiment, since we observe such segmentations consistently yield satisfactory performances. Each Elasticsearch index is created with five shards and zero replica on a single-node Elustersearch cluster deployed on a Microsoft Azure virtual machine with 12 cores and 112 GiB of RAM. 

%Our dataset $\mc B$ consists of binary codes generated from half a million images selected from Jet.com's furniture catalog. Specifically, each image is first embedded into a vector in $\reals^{1536}$ by taking the output from the penultimate layer (i.e., the last average pooling layer) of the pretrained \textsc{Inception-ResNet-V2} model \cite{szegedy2017inception}, and then hashed into compact binary codes in $\set{0,1}^m$ using iterative quantization (ITQ) \cite{gong2011iterative}, where $m \in \set{128, 256}$. Each Elasticsearch index is created with five shards and zero replica on a single-node Elustersearch cluster deployed on a Microsoft Azure virtual machine with 12 cores and 112 GiB of RAM. 

\paragraph{Evaluation.} We randomly select 1,000 binary codes from $\mc B$ to act as query codes $\bm q$.  For each $\bm q$, we compare the search latencies among all four methods with Hamming distance $r \in \set{5,10,15,20}$.

\renewcommand{\arraystretch}{1.75}
\begin{table}[h!]
	\begin{tabular}{|C{.8cm}|C{.9cm}|C{2.5cm}|C{2.5cm}|C{2.5cm}|C{2.5cm}|}
		\hline 
		$m$ & $r$ & {Term match} & {Bit operation} & {FENSHSES w/o prep. } & {FENSHSES}
		\\
		\hline  
		\hline
		\multirow{3}{*}{128} & 5 & 641.99 (\scriptsize 19.01) & 41.38 (\scriptsize 6.38) & 2.80 (\scriptsize 3.50) & 1.08 (\scriptsize 1.25) \\ 
		& 10& 638.20 (\scriptsize 16.65) & 42.24 (\scriptsize 7.39) & 7.40 (\scriptsize 5.07) & 3.62 (\scriptsize 1.54)\\ 
		& 15&  637.63 (\scriptsize 16.14) & 43.08 (\scriptsize 7.90) & 7.19 (\scriptsize 5.09) & 3.45 (\scriptsize 1.55)\\ 
		& 20& 638.41 (\scriptsize 17.41) & 42.65 (\scriptsize 7.59) & 15.51 (\scriptsize 5.88) & 9.51 (\scriptsize 2.18)\\ 
		\hline
		\multirow{3}{*}{256} & 5 & 1259.22 (\scriptsize 30.66) & 75.35 (\scriptsize 11.87) & 6.24 (\scriptsize 6.48) & 2.18 (\scriptsize 2.02)\\ 
		& 10& 1257.04 (\scriptsize 20.68) & 75.06 (\scriptsize 11.27) & 6.28 (\scriptsize 6.63) & 2.13 (\scriptsize 1.97)\\ 
		& 15& 1270.38 (\scriptsize 25.88) & 75.81 (\scriptsize 12.22) & 6.70 (\scriptsize 6.93) & 2.09 (\scriptsize 1.56) \\ 
		& 20& 1278.47 (\scriptsize 25.56) & 75.50 (\scriptsize 11.94) & 18.02 (\scriptsize 10.71) & 7.67 (\scriptsize 2.85) \\ 
		\hline
	\end{tabular}
	\caption{\textbf{Means and standard deviations (in brackets) of search latency (measured in ms) under different scenarios.} FENSHSES is dramatically faster than the term match approach, and all of the three techniques  involved in FENSHSES contribute substantially to this performance improvement. } \label{tab: search_latency}
\end{table}

\begin{figure}[h]
	\centering
	\includegraphics[width=.75\textwidth]{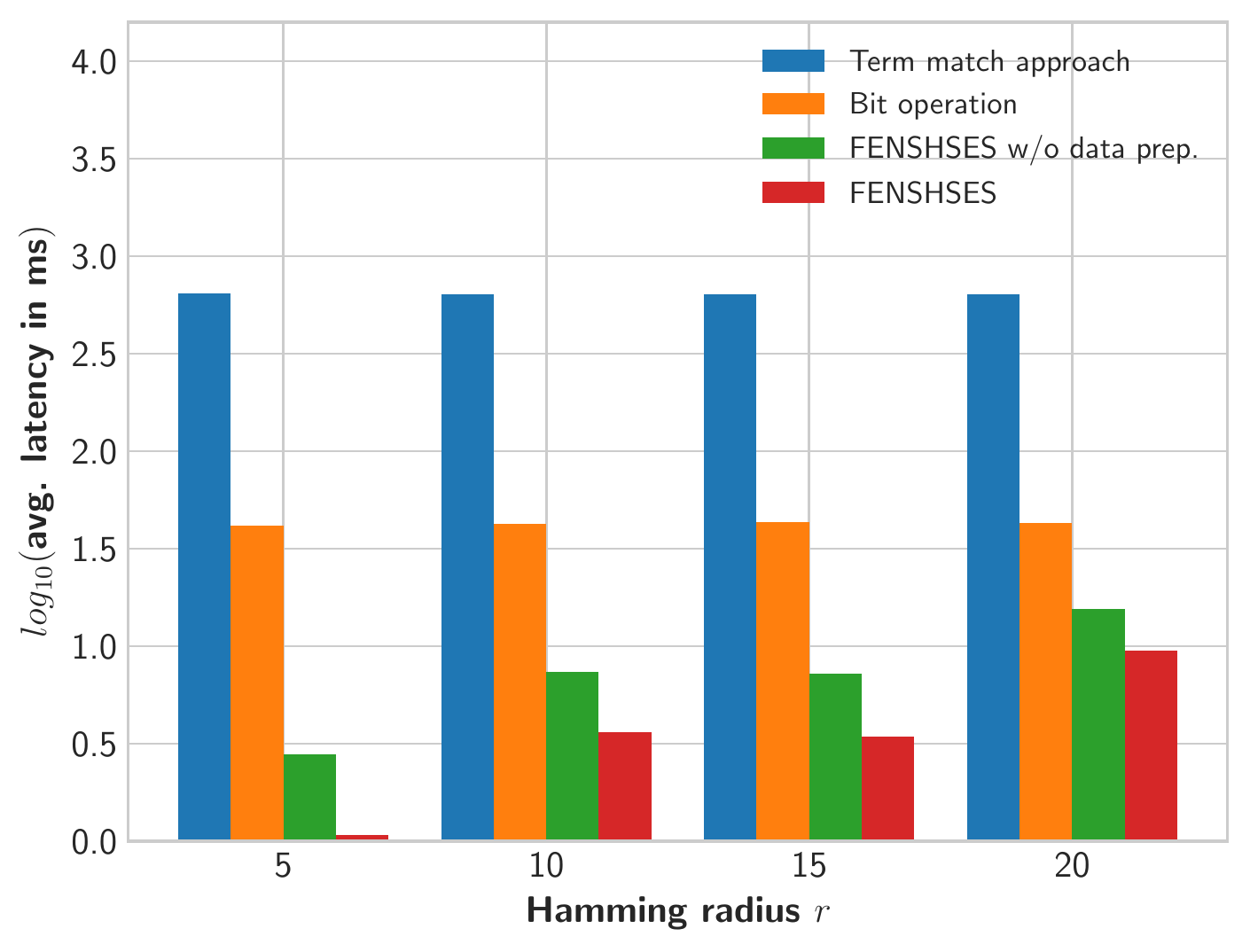}
	\caption{ \normalfont{Experimental results for 128-bit binary codes.}}
	\label{fig: 128-bit}
\end{figure}

\begin{figure}[h]
	\centering
	\includegraphics[width=.75\textwidth]{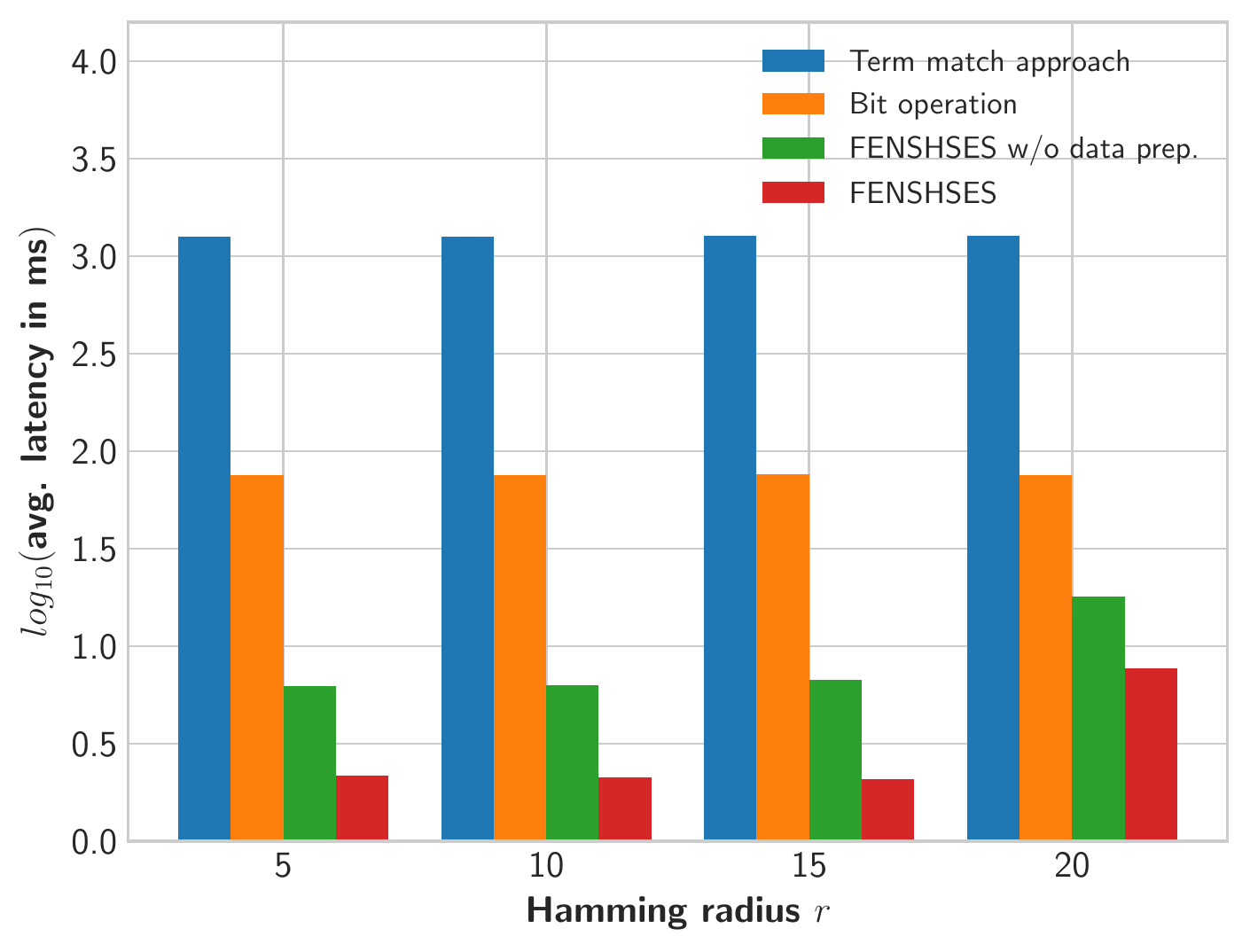}
	\caption{ \normalfont{Experimental results for 256-bit binary codes.}}
	\label{fig: 256-bit}
\end{figure}

\paragraph{Results.} 
As shown in Table \ref{tab: search_latency}, FENSHSES is much faster than the term match approach. To better visualize these speed-ups, we plot average search latencies under logarithmic transformation in Fig. \ref{fig: 128-bit} and  \ref{fig: 256-bit}. In the following, we address the contribution of each component of FENSHSES respectively.
\begin{itemize}
	\item By computing the Hamming distance using bit operation instead of term match, we consistently observe 
	 around sixteen times speedup over different $m$ and $r$. 
	\item The amount of speed-up introduced by sub-code filtering varies with the radius $r$. Specifically, as $\lfloor \frac{16 r}{s} \rfloor$ heavily influence the number of data points to be considered for Hamming distance computation (see \eqref{eqn: substring-filter}), for $r$'s with the same value of $\lfloor \frac{16 r}{s} \rfloor$, the search latencies of FENSHSES w/o prep. are quite similar. As $\lfloor \frac{16 r}{s} \rfloor$  becomes larger, the sub-code filtering technique will become less effective. In practice, since we  most likely care about nearest neighbors within a small radius, the sub-code filtering technique should be capable of greatly reducing the search latency. 
	\item By reshuffling binary codes to reduce their correlations within each sub-code group, the technique of data processing with permutation not only accelerates FENSHSES in terms of the average search latency, but also stabilizes its overall performance with much smaller standard deviation. 
	\item A comprehensive comparison between FENSHSES and FAISS \cite{johnson2017billion}  in terms of indexing speed, search latency and RAM consumption is also conducted in  \cite{mu2019cmp}, where FENSHSES demonstrates competitive performance. 

\end{itemize}

%
%As shown in Fig. \ref{fig: 128-bit} and  \ref{fig: 256-bit}, FENSHSES is dramatically faster than the term match baseline, and all of the three techniques involved in FENSHSES contribute substantially to this performance improvement. Specifically, the speed-ups achieved range from one hundred times  (for $r=20$) to six hundred times (for $r=5$). This difference in speedup is much expected, as the sub-code filtering technique will be most effective when $r$ is small. 

\section{Conclusion}
It has been recently demonstrated that NNS systems built upon full-text search engines are capable of effectively reducing main memory consumption, coherently supporting multi-model search and being well-prepared for production deployment. Motivated by these clear advantages, in this paper, we explore how to empower full-text search engines to efficiently find nearest neighbors in Hamming space. By revisiting bit operation, sub-code filtering and data preprocessing with permutation, we propose a novel approach to accomplish this task, which shown empirically to be substantially faster than the term match approach (the state-of-art one for nowadays full-text search engines to find nearest neighbors within binary codes). By implementing the proposed approach non-trivially on the Elasticsearch platform, we delivered a cutting-edge engineering solution called FENSHSES. In the future, we will also explore how to implement our approach efficiently on other full-text search engines (e.g., Solr and Sphinx).

\section*{Acknowledgment}
We are grateful to three anonymous reviewers for their helpful suggestions and comments that substantially improve the paper. We would also like to thank Aliasgar
Kutiyanawala for helping us fix a bug in an earlier version of JSON \ref{json: hmd64bit}, and thank 
Eliot P. Brenner for bringing the work \cite{rygl2017semantic} to our attention.

 \bibliographystyle{splncs04}
 \bibliography{vss}

%\bibliographystyle{ACM-Reference-Format}
%\bibliography{vss}
\end{document}